\documentclass[twoside,a4paper]{article}
\usepackage{fleqn,espcrc2}
\usepackage{graphicx}
\usepackage{latexsym}
\usepackage{euscript}
\newcommand{\ifig}[1]{\includegraphics[height=75mm,width=80mm]{#1}}
\newcommand{\bc}{\begin{center}}
\newcommand{\ec}{\end{center}}
\newcommand{\be}{\begin{equation}}
\newcommand{\ee}{\end{equation}}
\newcommand{\bsp}{\begin{split}}
\newcommand{\esp}{\end{split}}
\newcommand{\bea}{\begin{eqnarray}}
\newcommand{\eea}{\end{eqnarray}}
\newcommand{\ba}{\begin{eqnarray}}
\newcommand{\ea}{\end{eqnarray}}
\newcommand{\bas}{\begin{eqnarray*}}
\newcommand{\eas}{\end{eqnarray*}}

\newcommand{\simge}{\ \lower-
1.2pt\vbox{\hbox{\rlap{$>$}\lower5pt
\vbox{\hbox{$\sim$}}}}\ }

\newcommand{\AmS}{{\protect\the\textfont2
  A\kern-.1667em\lower.5ex\hbox{M}\kern-.125emS}}
\title{ 
 Quark and Gluon Propagators in Covariant Gauges
}
\author{L. Giusti\address{Centre de Physique Theorique
CNRS Luminy, Case 907
F-13288 Marseille Cedex 9 France
},
 M. L. Paciello\address{INFN, Sezione di Roma 1,
 P.le A. Moro 2, I-00185 Roma, Italy},
 S. Petrarca\thanks{Speaker at the Conference}$^{\rm b~}
 $\address{Dipartimento di Fisica, Universit\`a di Roma ``La
                     Sapienza''}%
        ,
B. Taglienti$^{\rm b}$
, N. Tantalo$^{\rm c}$
                     }

\begin{document}

\begin{abstract}
We present data for the gluon and quark propagators computed
in the standard lattice Landau's gauge and for three values 
of the covariant gauge-fixing parameter ($\lambda=0,8,16$).
Our results are obtained using the $SU(3)$ Wilson action
in the quenched approximation at $\beta=6.0$ and 
$V=16^3\times 32$.
\vspace{1pc}
\end{abstract}

\maketitle

In the last few years we have proposed a new 
lattice gauge-fixing algorithm~\cite{giusti}-\cite{covar1}
to study in a non perturbative framework the gauge dependence of 
the fundamental QCD correlators, i.e. gluon and quark propagators.
Since a na{\"\i}ve generalization of the 
standard Landau gauge-fixing functional is not 
possible~\cite{giusti},
we propose to fix a generic covariant gauge 
by minimizing a discretized version of 
the functional 
\be\label{cov11}
H_A[G]\equiv
\int d^4x\mbox{\rm Tr}\left[(\partial_{\mu}A^G_{\mu}-\Lambda)
(\partial_{\nu}A^G_{\nu}-\Lambda)\right]
\ee
where the $\Lambda$'s belong to the Lie algebra of the
group and are generated with a Gaussian distribution
$\exp[-1/\lambda \int d^4x \mbox{Tr}(\Lambda^2)]$
(for all notations and conventions see~\cite{rev},\cite{covar1}).  
We have implemented a discretization of 
the functional (\ref{cov11}) which is linear in any single 
local gauge transformation. This can be achieved with 
a careful manipulation of ${\cal O}(a)$ terms in the lattice 
functional definition~\cite{covar1}. 

We retrieved $221$ $SU(3)$ gauge configurations 
at $\beta=6.0$ and $V=16^3\times 32$~\cite{GaugeConn}
available from the repository at the ``Gauge Connection''
(http://qcd.nersc.gov). Covariant gauges corresponding 
to $\lambda=0, 8$ have been fixed for all of them, while 
the gauge corresponding to $\lambda=16$ has been imposed 
on $80$ configurations only. We have required always a 
gauge fixing quality factor  $\theta< 10^{-6}$.
The calculation of the gauge-fixing  rotation has been performed on the 
Boston University's Origin2000. 

Last year we presented in Bangalore~\cite{banga}
some preliminar results on the gluon
propagator in the $SU(3)$ quenched theory  
computed in the standard Landau and in the covariant gauge at two values
of the parameter $\lambda=0,8$. 
Here we present another set of data ($\lambda=16$) for the
gluon propagator and the quark
propagator, with the Wilson fermionic action, for three values of the
gauge parameter $\lambda=0,8,16$. 

\begin{figure}[h]
\bc
\ifig{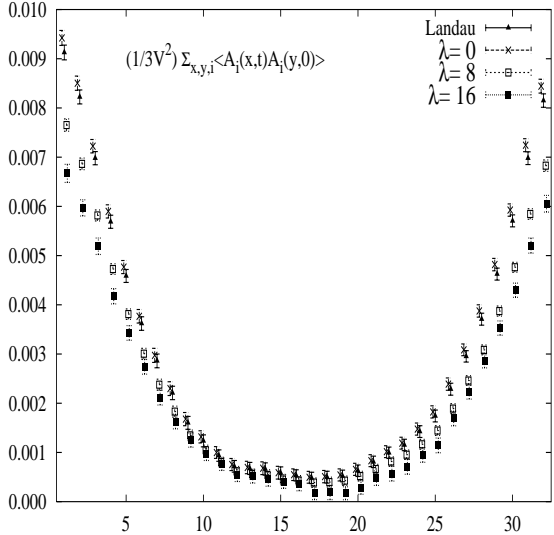}
\caption{\small{
Gluon propagator  as a function of~$t$.
 }}
\label{fig:gluo(t)}
\ec
\end{figure}

In Fig.~\ref{fig:gluo(t)} we show the behaviour of the 
the gluon propagator at zero spatial momentum
\be
\langle A_iA_i\rangle (t) \equiv \frac{1}{3 V^2}
 \sum_{i}\sum_{{\bf x},{\bf y}} Tr \langle  A_i({\bf x},t)A_i({\bf y},0)\rangle
\label{eq:AiAi}
\ee
as a function of $t$.
While the data for the standard Landau gauge 
and for $\lambda=0$ are compatible within errors,
the results for $\lambda=8,16$ show signals of a  
gauge dependence of the gluon propagator.

\begin{figure}[h]
\bc
\ifig{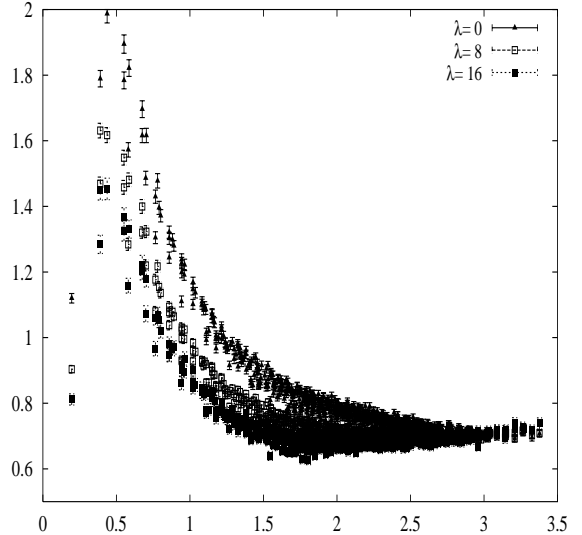}
\caption{\small{ Transverse part of the gluon propagator $q^2 D_T(q)$ in
covariant gauges as a function of $q$.
}}
\label{fig:tran}
\ec
\end{figure} 
\begin{figure}[h]
\bc
\ifig{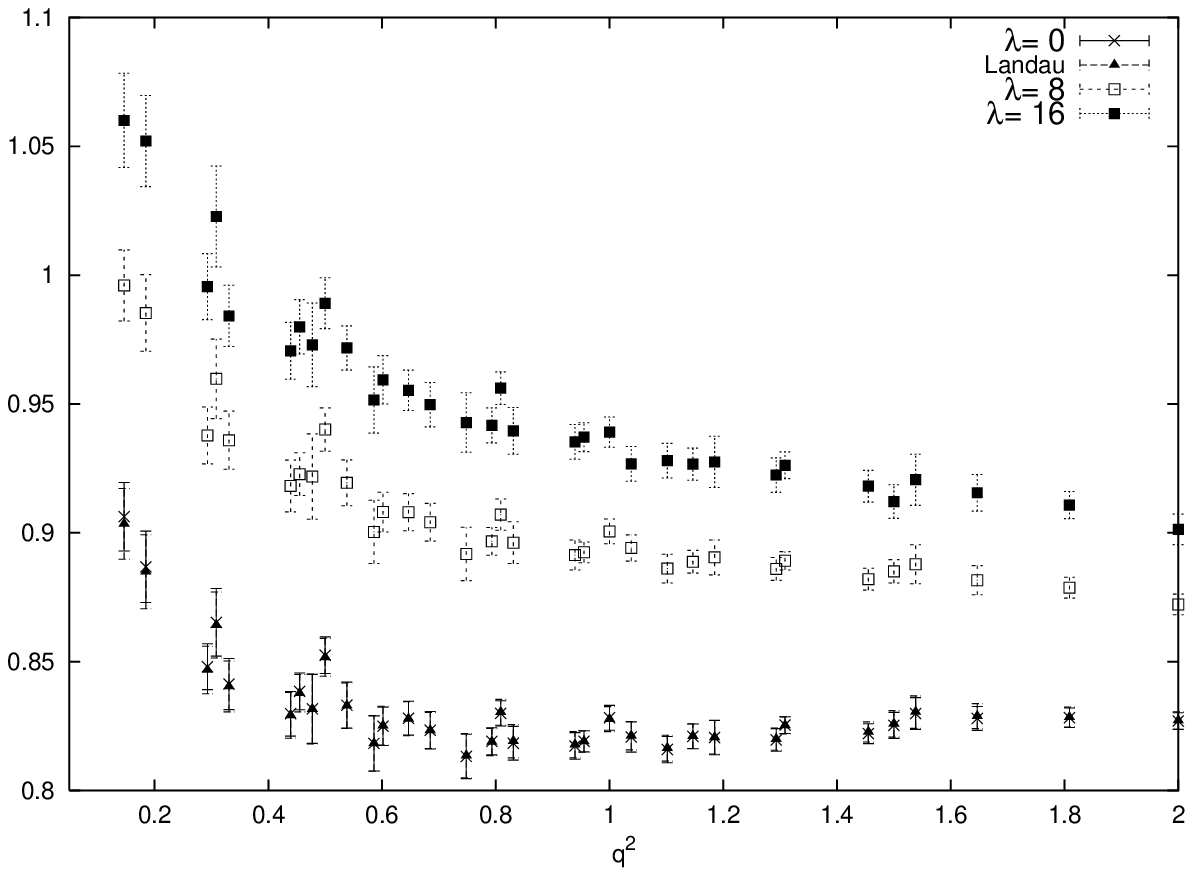}
\caption{\small{$\Sigma_1$ part of the quark propagator as a function of $q^2$ at 
$\lambda=0,8,16$ and $K=0.1515$.
}}
\label{fig:quark1}
\ec
\end{figure}
We have analyzed the data of the gluon propagator also in the Fourier space
by using the usual separation of transverse and longitudinal 
components, i.e. 
\be
D_{\mu\nu}(q)=(\delta_{\mu\nu} - {q_\mu q_\nu \over q^2})D_T(q)
+{q_\mu q_\nu \over q^2}{D_L(q) \over q^2}\; .
\label{eq:lontra}
\ee
In Fig.~\ref{fig:tran} we present the behaviour of the transverse
part $q^2 D_T(q)$ as a function of
$q=\sqrt{ q^2}$. The data have been averaged over
the $Z_3$ group~\cite{parri} and only values corresponding to 
$k_0={2 \pi \over 32} (0 \div 12)$ and
$k_i={2 \pi \over 16} (0 \div 6)$ are reported.
We see that at large $q^2$ the data at the three values of the gauge parameter
tend to coincide, while at low
$q^2$ they exhibit a clear gauge dependence. 
The data with higher $\lambda$ show a lower  peak
than those with $\lambda=0$, but the peak position remains fixed.
Note that in this region  physical effects caused by
  confinement are expected, see for example~\cite{reinhardt}. 

\begin{figure}[h]
\bc
\ifig{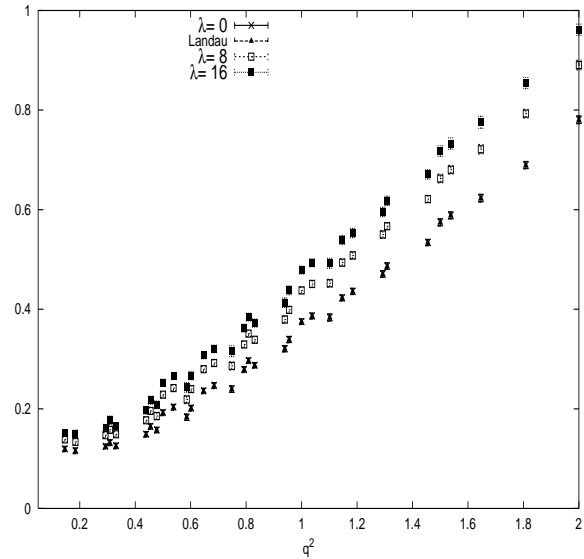}
\caption{\small{$\Sigma_2$ part of the quark propagator as a function of $q^2$ at 
$\lambda=0,8,16$ and $K=0.1515$. 
}}
\label{fig:quark2}
\ec
\end{figure}

The inverse of the quark propagator $S^{-1}(q)$, at $K=0.1515$,
 has been decomposed in 
the standard way
\be
S(q)^{-1}= i\not\!{q } \Sigma_1(q^2,m) + m \Sigma_2(q^2,m)\; .
\ee
In Fig.~\ref{fig:quark1} we report $\Sigma_1$ as a function of $q^2$.
At large $q^2$ in the Landau gauge, $\Sigma_1$ is expected to be a constant at 
leading order in perturbation theory. On the contrary,
in the covariant gauge it should decrease logarithmically
as a function of $q^2$~\cite{tantalo} with a slope which 
depends on $\lambda$. 
 The data can be fitted using the perturbation
expansion in order to extract the gauge fixing parameter
renormalization constant $Z_\lambda$. 
 The data favor a small value of the gauge parameter
renormalization constant: $Z_\lambda$$\simeq .35$ for $\lambda=8$ 
and  $Z_\lambda$ $\simeq .20$ for $\lambda=16$.
In order to complete the presentation of our results, we show
in Fig.~\ref{fig:quark2} the behaviour of $\Sigma_2$. It is well 
known~\cite{rapgiu} that in the Wilson fermionic regularization
this function is affected from large ${\cal O}(a)$ effects and
our data confirm this pathology in the covariant gauge too. 
In conclusion, our data show signals of gauge dependence 
in the gluon and quark propagator. We observe 
a small value of $Z_\lambda$ with a large dependence 
on $\lambda$ which seems to reduce the separation among
gauges with different $\lambda$. 

Our study implicitly
indicates that lattice Gribov copies
do not have a statistical significance 
in the quark and gluon propagators in  
the Landau gauge.
In fact Gribov copies are  averaged with 
different weights in the standard Landau gauge-fixing
and in the covariant gauge at $\lambda=0$. 
Nevertheless anything different turns out to be within 
our statistical errors.
 
We warmly thank Massimo Testa for many fruitful discussions.
We thank the Center for Computational Science of Boston 
University where the gauge fixing was performed.
We also gratefully acknowledge the use of the gauge configurations 
produced by the authors in Ref.~\cite{GaugeConn}.


\end{document}